# Ultrafast MHz-rate burst-mode pump-probe laser for the FLASH FEL facility based on nonlinear compression of ps-level pulses from an Yb-amplifier chain

*Marcus Seidel,[1] Federico Pressacco,[1] Oender Akcaalan,[1] Thomas Binhammer,[2] John Darvill,[1] Nagitha Ekanayake,[1] Maik Frede,[2] Uwe Grosse-Wortmann,[1] Michael Heber,[1] Christoph M. Heyl,[1,3,4] Dmytro Kutnyakhov,[1] Chen Li,[1] Christian Mohr,[1] Jost Müller,[1] Oliver Puncken,[2] Harald Redlin,[1] Nora Schirmel,[1] Sebastian Schulz,[1] Angad Swiderski,[1] Hamed Tavakol,[1] Henrik Tünnermann,[1] Caterina Vidoli,[1] Lukas Wenthaus,[1] Nils Wind,[5,6] Lutz Winkelmann,[1] Bastian Manschwetus,[1] and Ingmar Hartl[1]*

E-mail: marcus.seidel@desy.de federico.pressacco@desy.de

[1] Deutsches Elektronen-Synchrotron DESY, Notkestraße 85, 22607 Hamburg, Germany

[2] neoLASE GmbH, Hollerithallee 17, 30419 Hannover, Germany

[3] Helmholtz-Institute Jena, Fröbelstieg 3, 07743 Jena, Germany

[4] GSI Helmholtzzentrum für Schwerionenforschung GmbH, Darmstadt, Germany

[5] Physics Department University of Hamburg and Centre for Free-Electron Laser Science (CFEL), 22761 Hamburg, Germany

[6] Ruprecht Haensel Laboratory, Deutsches Elektronen-Synchrotron DESY, D-22607 Hamburg, Germany

*Keywords:* ultrafast lasers, high-power lasers, pulse compression, spectral broadening, lasers for facilities, synchronization, pump-probe

## Abstract

The Free-Electron Laser (FEL) FLASH offers the worldwide still unique capability to study ultrafast processes with high-flux, high-repetition rate XUV and soft X-ray pulses. The vast majority of experiments at FLASH are of pump-probe type. Many of them rely on optical ultrafast lasers. Here, a novel FEL facility laser is reported which combines high average power output from Yb:YAG amplifiers with spectral broadening in a Herriott-type multi-pass cell and subsequent pulse compression to sub-100 fs durations. Compared to other facility lasers employing optical parametric amplification, the new system comes with significantly improved noise figures, compactness, simplicity and power efficiency. Like FLASH, the optical laser operates with 10 Hz burst repetition rate. The bursts consist of 800 μs long trains of up to 800 ultrashort pulses being synchronized to the FEL with femtosecond precision. In the experimental chamber, pulses with up to 50 μJ energy, 60 fs FWHM duration and 1 MHz rate at 1.03 μm wavelength are available and can be adjusted by computer-control. Moreover, nonlinear polarization rotation is implemented to improve laser pulse contrast. First cross-correlation measurements with the FEL at the plane-grating monochromator photon beamline are demonstrated, exhibiting the suitability of the laser for user experiments at FLASH.

## 1. Introduction

The superconducting Free-Electron Laser (FEL) FLASH provides ultrashort, extremely powerful pulses in the XUV and soft X-ray spectral range (1.5 nm to 50 nm) at the highest repetition rates worldwide. Since more than 80% of the experiments at FLASH are time-resolved pump-probe experiments, femtosecond optical pulses constitute a vital cornerstone of contemporary FEL experiments. In particular, the plane-grating (PG) monochromator photon beamline at FLASH[1–3] is constantly in high demand and was booked in each of the last four years for more than 50% of all science experiments using pump-probe lasers. The beamline serves predominantly the condensed matter science community using methods such as time-resolved photoelectron, X-ray absorption and X-ray emission spectroscopy which typically require sub-100 fs pulses in the near-infrared spectral region for non-resonant sample excitation. To enable FEL users to fully exploit their limited beam time, facility lasers must be operational 24/7. At FLASH, the availability of the optical lasers for pump-probe experiments was above 95 % of the requested time over the past 3 years. To achieve such long-term performance, the lasers at FLASH host various online diagnostics and are to a large extend remotely controllable. Ultimately, excellent passive stability is desired which calls for simple and compact laser systems.

During the past ten years, two optical lasers were available at the FLASH1 beamlines:[4] First, a Ti:sapphire laser with 10 Hz repetition rate providing mJ-level pulse energies. Second, a μJ-level laser based on optical parametric chirped pulse amplification (OPCPA). This source emitted bursts adapted to the FEL pulse sequence (cf. **Figure 1** top panel). FLASH operates in burst-mode with 10 Hz repetition rate. The 800 μs long bursts are again comprised of femtosecond pulse trains with 1 MHz repetition rate. Laser emission adapted to this pulse sequence is hence ideal for applications. At the PG beamlines, the requested pulse energies are moderate, that is typically on the 1 µJ order, but the requested intra-burst repetition rates are high, usually between 100 kHz and 1 MHz. This allows to take full advantage of the unique pulse rates of the FEL which enables photon-hungry applications at the beamline. The previously used, complex OPCPA system was decommissioned in fall 2020 and has now been replaced by a much simpler laser system which is reported here. It is the first optical laser in FEL beamline user operation which relies on the concept of nonlinear spectral broadening in Herriott-type multipass cells (MPC).[5,6] The method enables compression of high-power ps-level pulses from Yb-based lasers to sub-100 fs duration with a compact setup, very good intra-burst pulse energy flatness and excellent burst-to-burst energy stability. The reported system provides multi-µJ pulses at 1030 nm with durations down to 60 fs. It furthermore contains a pulse shaping unit for improved pulse contrast in a time window of interest for user experiments. A burst energy stability of 0.5 % rms over 24 hours is demonstrated which is an order of magnitude better than the stability of the previous burst-mode laser.[4] Finally, extensive diagnostics combined with field-programmable gate arrays (FPGA) and programmable logic controllers (PLC) offer FEL users to monitor and control the laser parameters by means of DESY's accelerator control system. As a result, the laser described here provides a stable and reliable new workhorse for users of the PG beamlines at FLASH. It will be an essential ingredient for upcoming cutting-edge FEL experiments.

## 2. Laser setup

The laser is installed in a 12 m$^2$ large so-called modular optical delivery station in the FLASH1 experimental hall. For the sake of clarity, the laser setup has been subdivided into five parts which are highlighted by different background colors in **Figure 1**.

### 2.1. Burst Generation and Amplification

In section A, a soliton mode-locked Fabry-Pérot-type fiber oscillator generates ultrashort pulses.[7] Those are amplified in three consecutive single-mode Yb-doped fiber amplifiers (YDFAs). Subsequently, the ~ 15 nJ pulses from the fiber frontend are amplified up to 200 µJ at the plateau of the laser bursts in four Yb:YAG end-pumped rods, that is about 200 W intra-burst average power. The solid-state amplifier design is adapted from a four stage Nd:YVO$_4$ amplifier implemented at the European XFEL photocathode laser.[8] The pulse energies can only be reached by chirped pulse amplification (CPA). Consequently, the pulses are stretched by a 200 m long fiber behind YDFA 1 and recompressed to about 900 fs by a 4-pass single-grating compressor. Section A furthermore contains 3 acousto-optic modulators (AOMs). They down-pick the pulse repetition rate and control the laser pulse intensities along the bursts as visualized in the top panel of **Figure 1**. Iterative learning control[9] is applied to achieve constant pulse energies over the burst by means of feedback to the AOM 3 modulation port. AOM 3 also sets the pulse repetition rate within the 800 μs long burst according to user demands. The maximum of 1 MHz matches the FEL pulse rate. The fiber frontend has a second output behind YDFA 4. Short pulses (170 fs) at the oscillator's repetition rate (54 MHz) are generated to synchronize the optical pulses with the FEL bursts. The footprint of section A is only about 2 m x 0.7 m. More details are provided in supplement 1.

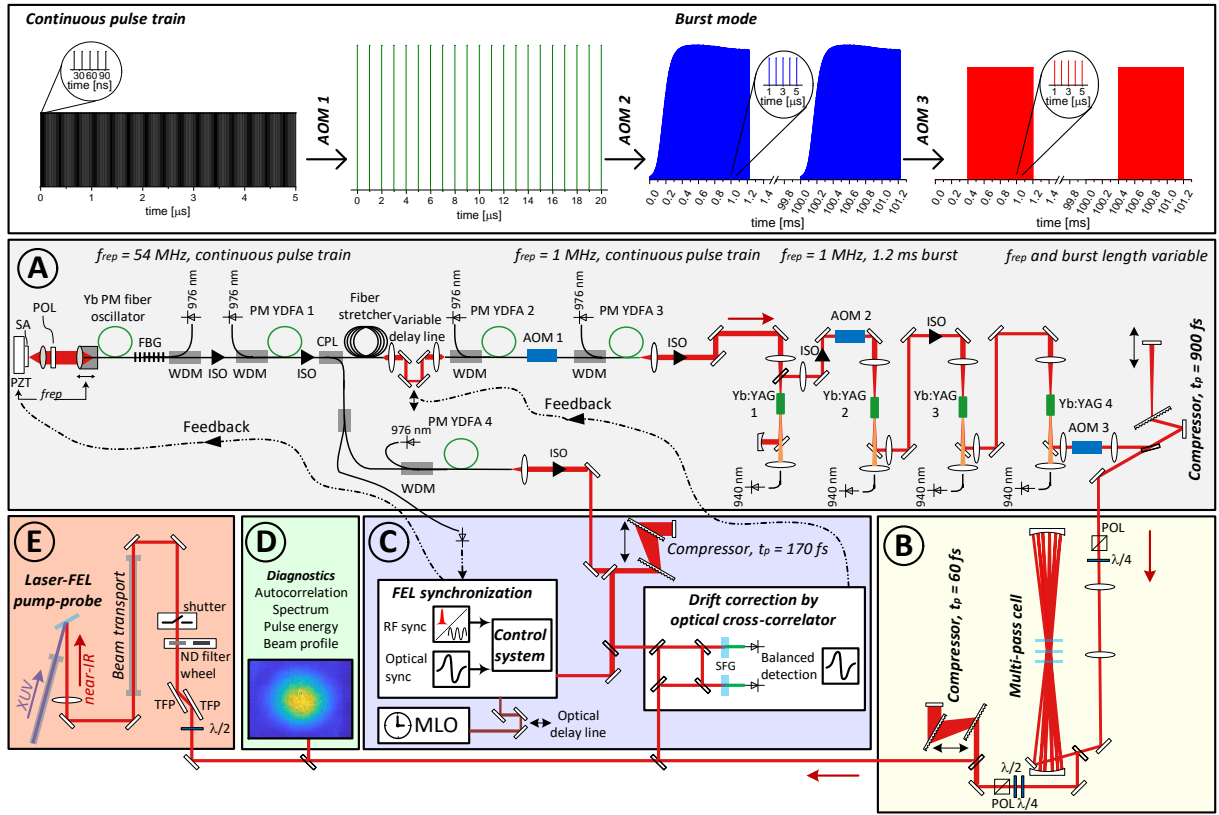

**Figure 1.** The top panel shows the pulse trains and burst shapes, respectively. The other panels give an overview of the laser setup consisting of five sections: A - Burst generation and amplification section consisting of a fiber oscillator with controlled cavity length, four fiber preamplifiers, four bulk Yb:YAG main amplifiers, three AOMs for burst preparation and shaping, a delay line for drift control and a fiber stretcher as well as a grating compressor for dispersion control. B - Pulse compression section consisting of an MPC, polarization optics and a grating compressor. C – Synchronization and timing section consisting of two balanced cross-correlators; D – Laser diagnostics section consisting of near- (shown beam profile) and far-field cameras, a spectrometer, an autocorrelator and an energy meter, E – laser delivery section consisting of a variable attenuator and the beam transport unit to the PG beam line. Black solid lines denote fiber, red / orange solid lines free path optics, black dashed lines electrical signals. AOM – acousto-optic modulator, CPL – coupler, FBG – fiber Bragg grating, $f_{rep}$ – repetition rate, ISO – optical isolator, MLO – main laser oscillator, ND – neutral density, POL – polarizer, PM – polarization maintaining, SA – saturable absorber, SFG – sum frequency generation, TFP – thin-film polarizer, $t_p$ – pulse duration, WDM – wavelength division multiplexer, YDFA – Ytterbium-doped fiber amplifier

## 2.2. Pulse compression

To increase the temporal resolution of the FEL pump-probe experiments, spectral broadening in a Herriott-type MPC and consequent pulse shortening in a grating compressor was used (Section B of **Figure 1**). Whereas the majority of reported experiments rely on a single Kerr medium within the MPC, a hybrid multi-pass multi-plate approach was implemented.[10] By this method the hitherto published pulse compression factors from a single bulk MPC[11,12] were clearly surpassed. Three 1 mm thin anti-reflection (AR) coated silica plates with 3 cm spacing were placed in the center of an about 350 mm long MPC. For maximized spectral broadening to a 55 fs Fourier transform limit, about 80 % of the 115 μJ input pulse energy was transmitted through the output polarizer after 31 roundtrips. The pulses were compressed by a record-high factor of more than 15 with a motor-controlled double-pass grating stage. **Figure 2**a and b show the retrieval results of a scanning frequency-resolved optical gating (FROG) measurement. The retrieved spectrum agrees well with the one measured in parallel with a commercial grating spectrometer, indicating the reliability of the FROG result. Moreover, high spectral homogeneity over the beam profile (**Figure S2**) was obtained which is in good agreement with previous bulk-MPC experiments.[13,14] The pulses were nearly Fourier-transform limited but clearly show side lobes, owing to the modulated spectrum. **Figure 2**c displays the compression quality in terms of pulse energy in the main peak and amplitude of the dominant side peak. The quantities were measured at different intra-burst delays. This was possible by means of the modulation capabilities of AOM 3 which cut out single pulses from the burst. The relative pedestal amplitude was considered as the most critical parameter for FEL experiments. It amounts to about 10 % of the main peak and is therefore comparable to the pulse-energy fluctuations in the FEL burst.[15] The variations of the pulse shapes over the burst stemmed from the transient thermal lenses which are described in supplement 1.2. It is remarkable that the 30 % waist area variation measured in the $M^2$-meter results in only about 3.5 fs pulse duration variation over the burst (**Figure 2**d) which hardly affects the temporal resolution of the pump-probe-experiments at the PG beamline (section 3).

A 24-hour measurement was taken to investigate the stability of the nonlinearly broadened spectra (**Figure 2**f). Hardly any fluctuations are visible by eye. To quantify the stability, the Fourier transform limit within a 30 dB dynamic range was evaluated for each recorded spectrum, resulting in a standard deviation of only 0.4 fs at a 55.5 fs mean transform limit. This is on the sub-percent level of the temporal resolution attained by optical pulse - XUV FEL cross-correlation measurements.

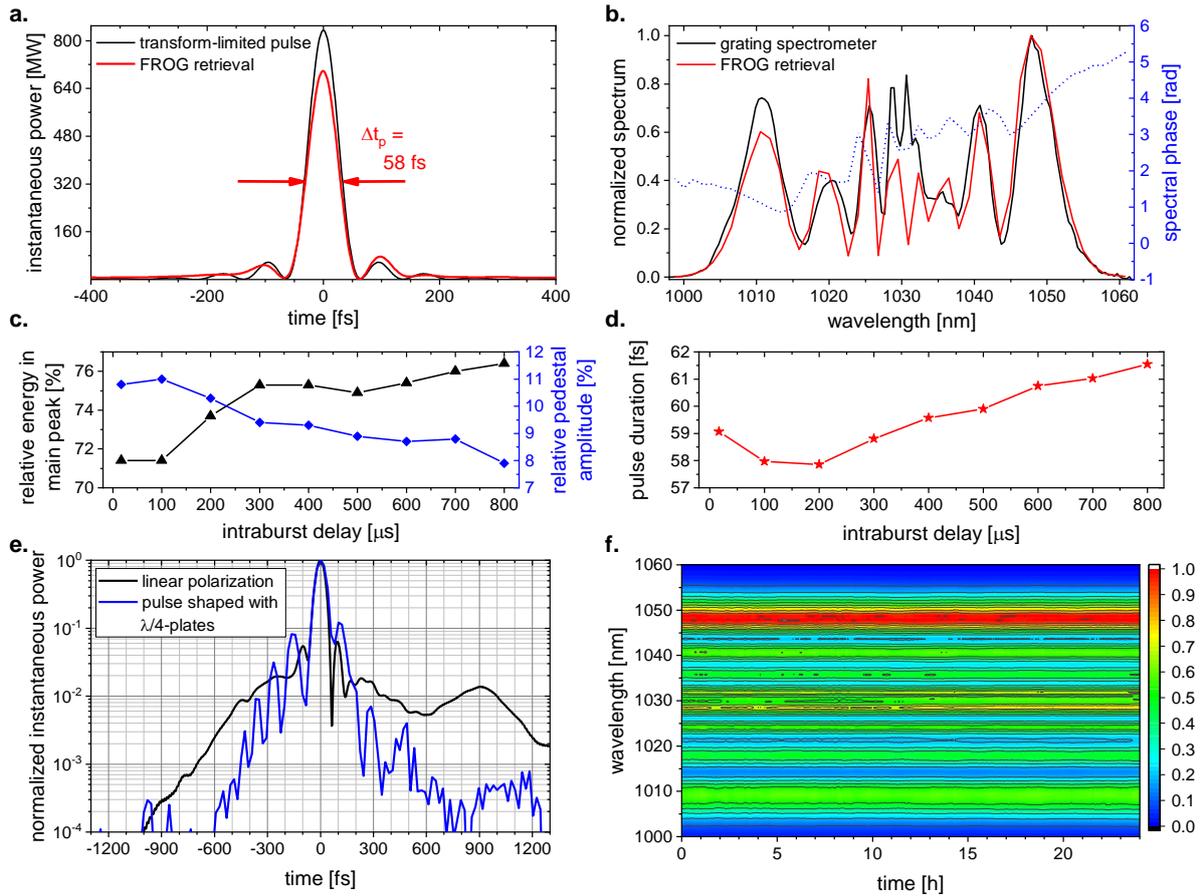

**Figure 2. a.** Retrieved FROG trace for 100 µs intra-burst delay and comparison to Fourier-transformed pulse derived from the retrieved spectrum shown in **b.** The spectral phase and a comparison to spectrum directly measured with a grating spectrograph are shown as well. **c.** Compression quality evaluated in terms of relative energy in the main peak defined by the minima around ± 60 fs and in terms of relative pedestal amplitude, that is peak power of the pedestal around 100 fs divided by the peak power of the main pulse. **d.** Varying pulse durations over burst length. The spread is about 3.5 fs, that is less than 7 % of the mean. **e.** Retrieved pulse from subplot **a** shown in logarithmic scale (black line) and compared to retrieved pulse after introducing nonlinear polarization rotation in the MPC. **f.** Logged normalized spectra measured over 24 hours every 10 s. The mean Fourier transform limit is 55.5 fs, its standard deviation 0.4 fs.

To suppress the pulse pedestals which inherently emerge from the self-phase modulated spectra, additional quarter-wave plates were placed at the entrance and the exit of the MPC. By introducing a slight polarization ellipticity with the entrance waveplate, the output polarization becomes intensity-dependent. Consequently, the polarizing beam splitter at exit of the MPC served as an artificial saturable absorber. This technique, called nonlinear polarization ellipse rotation, was previously used for pulse cleaning in fiber, single-pass nonlinear media and multi-pass geometries[16–18] but was for the first time directly integrated in an MPC-based spectral broadening unit. Pulse cleaning was introduced to suppress the post-pulse delayed by 900 fs from the main peak (**Figure 2**e, black solid line) because of its signature in a cross-correlation measurements with the FEL (supplement S6). Employing nonlinear polarization ellipse rotation suppressed the satellite pulse by more than an order of magnitude (**Figure 2**e, blue solid line). To accomplish this, both waveplates were manually adjusted such that the modulations of the spectrum after the MPC were minimized. The used configuration sufficed to quench the spurious signal emerging from the satellite pulse in FEL cross-correlation measurements (section 3). The introduction of an artificial saturable absorber by the ellipse rotation method reduced the mean pulse energy at the diagnostics section by 39 % to 35 µJ and increased the pulse width to about 70 fs owing to the polarization dependence of the nonlinear refractive index. Both drawbacks are however irrelevant for most of the user experiments which run at a few µJ pulse energy and with > 100 fs temporal resolution owing to the XUV-pulse stretching by the FEL beamline monochromator.

The footprint of section B is about 0.3 m x 0.9 m. Consequently, the whole pulse generation and shortening unit covers only an area of less than 2.5 m x 1 m, and is thus considerably more compact than the previous OPCPA laser system.

### 2.3. Synchronization and timing

A major task to ensure high time resolution of FEL-optical laser experiments is to synchronize both sources while providing precise control over their relative arrival time at the experiment. Therefore, the laser system includes two balanced optical cross-correlators[19,20] which are located in section C of the setup in **Figure 1** and explained in more detail in supplement 5. One cross-correlator is used for locking the fiber oscillator repetition rate to a main laser oscillator (MLO) operating at 1550 nm central wavelength. This facility-wide timing reference is distributed via length-stabilized fibers.[20] To minimize the cross-correlator's error signal a piezo-actuated end-mirror of the laser cavity is controlled to adjust the oscillator repetition rate. The control electronics are based on the MicroTCA4 platform developed at DESY[21] and a programmable FPGA hosting fast feedback loops. Since the capture range of the optical synchronization method is limited to about 400 fs, the fiber oscillator repetition rate is pre-stabilized by conventional RF phase-locking.[20] The 54 MHz pulse train from the auxiliary fiber frontend output is used in both cross-correlators. Whereas it is stabilized in the first cross-correlator, it is used as reference in the second correlator measuring slow timing drifts of the pulses coming from the MPC. A motor-controlled translation stage located in the fiber front-end serves as variable delay line to compensate for the measured drifts. The residual timing instability was logged over a 24-hour period during a user campaign (**Figure 3**a). The fast jitter derived from the 54 MHz pulse train was on average 25.1 fs within a 10 Hz to 1 MHz bandwidth. Only four out of > 85000 logged data points exhibited more than 30 fs timing-offset. In addition, more than 99.5 % of the slow drifts measured in the second cross-correlator were within a ±10 fs range and never exceeded a 20 fs absolute timing offset. Consequently, the total jitter between pump-probe laser and optical main oscillator is clearly below the > 100 fs temporal resolution achievable at the monochromator beamlines and thus hardly contributes to the overall resolution (**Figure 4**, **Figure S**).

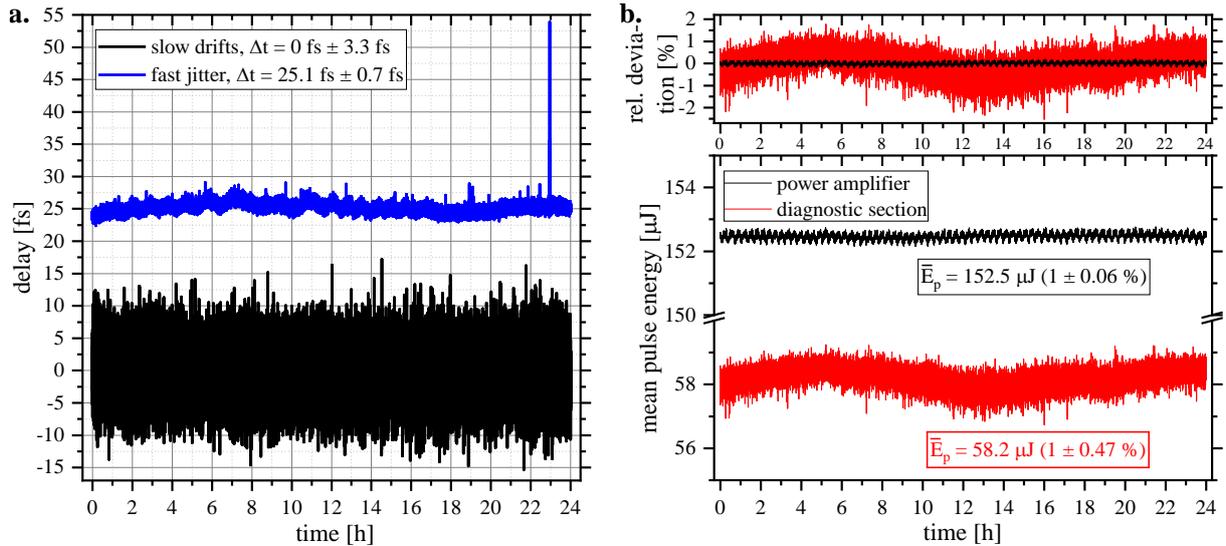

**Figure 3. a.** Pulse arrival time delay measurements in both balanced cross-correlators recorded in parallel to the energy stability in **b**. The blue line shows the fast jitter derived from the 54 MHz pulse trains. The black line shows the in-loop drift measurement of the 10 Hz laser bursts after nonlinear pulse compression. **b.** Energy stability measured over one day of user beamtime with a 1 Hz rate by photodiodes behind the main amplifier (black solid lines) and in the diagnostics section (red solid lines). The lower panel shows the absolute mean pulse energies inside the laser bursts, the upper panel shows the relative deviations from the mean value. The Yb:YAG amplifier is highly stable with < 0.1 % standard deviation, the MPC output shows also a low relative standard deviation of less than 0.5 %.

## 2.4. Laser stability

Section D of **Figure 1** hosts various diagnostic tools to measure pulse spectra, energies, durations and beam profile as well as position. The burst energy stability is continuously logged with a 1 Hz rate by the accelerator control system. **Figure 3**b shows the logged pulse mean energies at the diagnostics section. The energies are averaged over all pulses of the bursts. The data was recorded in parallel to the synchronization data shown in **Figure 3**a. With less than 0.1 % relative standard deviation, the solid-state amplifier exhibits a highly stable output. The burst energy fluctuations behind the MPC are higher but still below 0.5 % which is an order of magnitude better than the reported value for the previously used OPCPA system.[4] In parallel, beam displacement and pointing was analyzed in the same one-day measurement period at the diagnostics section. The standard deviations of the positions were 4.4 µm and 6.9 µm in x- and y-direction, respectively. This corresponds to 0.5 % of the $1/e^2$-beam radius in x- and 1.0 % in y-direction. The pointing standard deviations were 2.0 µrad and 2.6 µrad in x- and y-direction, respectively. The measurement data is shown in supplementary **Figure S3**. Users have not observed any problems with pointing during the first campaigns with the laser. It is to note that at present, the long-term stability strongly depends on the temperature and humidity stability of the FLASH experimental hall where the laser is installed without distinct air conditioning. Supplement S4.2 shows the mutual dependence of temperature, humidity and mean pulse energy at the diagnostics section.

### 2.4. Automation and controls

In order to rapidly adjust the laser settings to the user needs, to prevent drifts or damages, to minimize downtime and to continuously record pulse and beam parameters, the laser system is to a large degree remotely controllable and uses several automation routines. For this purpose, it has been integrated into the FLASH facility control system.[22,23]

The pulse trains are monitored by means of InGaAs photodiodes behind the oscillator, each fiber and solid-state amplifier, the fiber stretcher, the AOMs 1 and 3, at the compressor and MPC inputs, at the diagnostics section and the incoupling to the beamline. Pulse spectrum and autocorrelation as well as the near- and far-field beam profiles at the stabilization units behind the main amplifier and in the diagnostics section are continuously recorded. The MPC is equipped with three additional cameras for input- and output-beam profile and scattering light monitoring. If one of the Kerr media gets damaged, it can be replaced remotely by a slider which has however not been necessary, yet. The main amplifier Yb:YAG crystals are protected by an FPGA-based system which immediately interrupts the trigger to the pump diodes' power supply if a single seed pulse is missing in the burst or if the pulse energy drops below a set threshold value. The triggers of all photodiodes and measurement devices can be remotely controlled by laser experts who can also turn on and off all stages of the laser system remotely and synchronize it to the main laser oscillator. Moreover, the laser bursts after AOM 3 can be arbitrarily shaped and the intra-burst pulse repetition rate can be set (supplement S1.2). Finally, the pulse entering the MPC can be pre-chirped by the motorized grating stage in front of it.

Additionally, users of the PG beamline have several control options that fulfill their most common requests. First, the final section E of **Figure 1** contains a motor-controlled shutter and a variable attenuator unit. The pulse energies are adjusted by set of reflective neutral density filters and by a rotatable waveplate in front of two thin-film polarizers, resulting in 57.5 dB dynamic range. The burst energies can be directly measured by a pyroelectric energy meter addressable through a motorized flip mirror. Secondly, a high-precision, low drift translation stage is implemented for delaying the 1.55 μm reference beam (section C in **Figure 1**). It is used to control the relative delay between the FEL and the optical pulses. An automated routine, which shifts the overall laser timing up to several milliseconds, initiates the temporal overlap between pump and probe pulses. Thirdly, users can vary the duration of the optical pulses by changing the step motor-controlled grating separation in the compressor behind the MPC. Finally, the polarization of the light and the position of the beam in the experimental chamber is adjustable.

### 3. Free-electron laser pump-probe experiment

**Figure 4** shows the results of a proof-of-concept pump-probe experiment at the PG beamline. Optical and free electron lasers were overlapped at the surface of a tungsten sample to measure their cross-correlation signal.[24,25] The HEXTOF detection scheme[15] was used to collect the photoelectron yield in dependence of free electron kinetic energy and FEL-optical pulse delay. **Figure 4a** resolves the vicinity of the W(110) 4f core-level binding energies at $E_{b,1} \approx -31.3$ eV and $E_{b,2} \approx -33.4$ eV. Separated by multiples of the optical photon energy $h\nu_{nIR} = 1.2$ eV, transient dressed-states are formed symmetrically around the core-levels.[26,27] Consequently, integration over the energy range of 1.25 eV around $E_{b,1} + h\nu_{nIR}$ yields the FEL-optical laser cross-correlation signal.[28] **Figure 4b** shows the normalized data of the cross-correlation measurement and compares it to numerical calculations applying the equation

$$I_{XC}(\tau) \propto \int dt_0 P(t_0) \int dt\, I_{FEL}(t, t_0) I_{nIR}(t + \tau) \tag{1}$$

where the normal distribution $P(t_0) = \left(2\pi\sigma_{t_0}^2\right)^{-1/2} \exp\{-t_0^2/(2\sigma_{t_0}^2)\}$ with $\sigma_{t_0} = 30$ fs accounts for timing jitter, $I_{FEL}(t, t_0)$ is the FEL pulse intensity and $I_{nIR}(t + \tau)$ is the retrieved FROG pulse from **Figure 2e**. The FEL pulse was approximated by a Gaussian $I_{FEL}(t, t_0) \propto \exp\{-4\ln2(t - t_0)^2/\Delta t_p^2\}$ where the full-width half-maximum (FWHM) duration $\Delta t_p$ was derived from electron bunch streaking experiments[29] and the plane grating illumination (methods). Since the electron bunch distribution was not Gaussian, this is a coarse estimation. Nevertheless, the agreement between the measured data and the computed curve is good. The

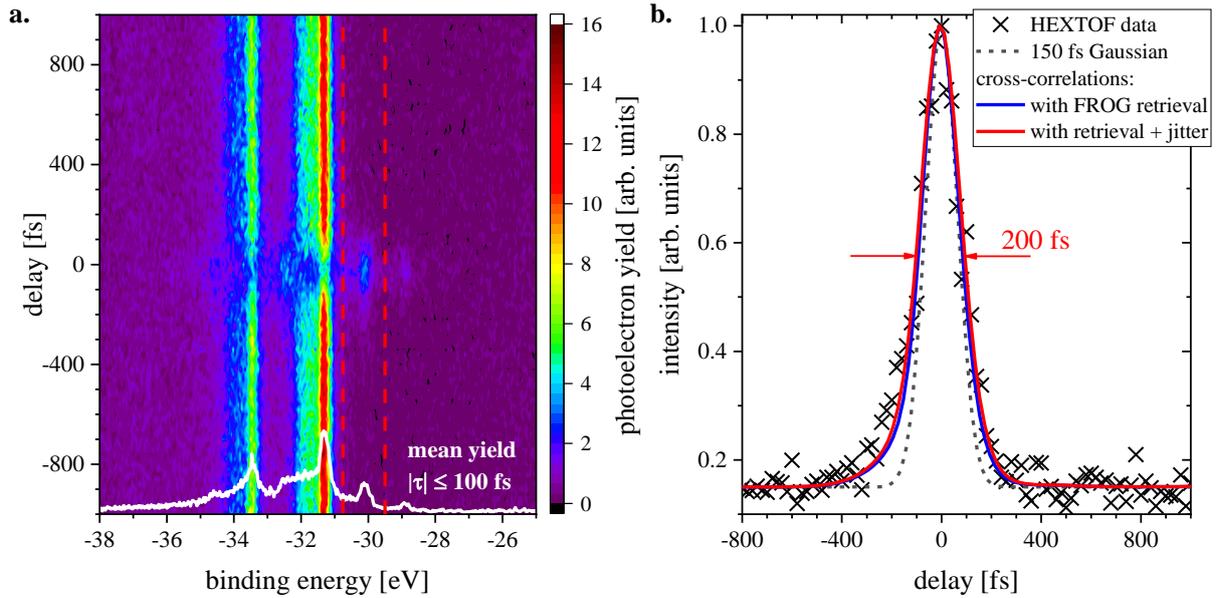

**Figure 4 a.** Measured photoelectron spectra of the W(110) 4f core level as function of time delay between optical laser and FEL ($h\nu_{FEL} = 112$ eV $\approx 93 \times h\nu_{nIR}$). At delays larger than $|\pm 600$ fs|, the binding energies virtually represent the steady state of the 4f core level. On contrary, for delays within $\pm 100$ fs (white solid line), optical photon dressed states emerge separated by multiples of $\pm h\nu_{nIR}$. They are visible up to the 2$^{nd}$ order near -28.9 eV. Clearly most prominent is the single-photon side-band at $E_{b,1} + h\nu_{nIR} \approx -30.1$ eV. The integrated photoelectron yield in a range of 1.25 eV around this band (red dashed lines) yields the cross-correlation measurement of **b.** The integrated data is represented by the black crosses. Moreover, the cross-correlation was modeled by Eq. (1) based on the FROG retrieval of the optical pulse. The blue solid line shows the expected cross-correlation excluding timing jitter between the FEL and optical pulses. The red line additionally considers jitter of 30 fs rms. For comparison, the estimated FEL pulse is shown (grey dashed line) which mainly determines the temporal resolution of the experiment. No side lobes emerge in the cross-correlation.

200 fs width of the cross-correlation is mainly determined by the 150 fs FEL pulse duration. The optical pulse adds about 45 fs to it. The impact of the 30 fs rms jitter is very small. In order to study the best temporal resolution at the PG beamline, shorter FEL pulses must be generated. This will be subject to an upcoming study. The cross-correlation trace exhibits broader wings than a Gaussian of comparable width. This originates at least to some degree from the pedestals of the optical pulse at $\pm 100$ fs (**Figure 2**e). Those are smeared out by the longer FEL pulses. Contrary to the cross-correlation measurement with the optical pulses shown in **Figure 2**a (supplement 6), no satellite pulses are visible in the cross-correlation.

## 4. Discussion

Efficient nonlinear pulse compression allows to take advantage of high-power Yb-ion based lasers without compromising temporal resolution. This promises a significant boost of on-target laser fluence in FEL pump-probe experiments, and thus constitutes a crucial building block of the FLASH 2020+ upgrade which targets THz to UV spectral coverage by optical lasers.[30] Precedent experiments with MPCs at FLASH have already demonstrated the potential of nonlinear pulse compression at FEL facilities.[31,32] Here, the first laser system is reported which fully relies on the spectral broadening concept and has been employed in FEL user experiments.[33] To date, PG beamline users have given positive feedback to the novel pump laser system.

In contrast to the previous OPCPA concept,[4] the system reported here exhibits various advantages: First, the order-of-magnitude energy stability improvement implies a clear reduction of measurement noise and reduced need for averaging, respectively. Pulse-to-pulse energy fluctuations are secondly also improved by active burst flattening through the last AOM in the amplifier unit. Such flattening was not possible in the OPCPA setup due to the a priori sub-100 fs bandwidth of the parametric amplifier seed. Third, the new system delivers up to 800 pulses per burst corresponding to the number of FEL pulses arriving at the experimental chamber. The previous system could lately deliver only 400 pulses per burst in best case[15]. The spectral broadening-based system can hence improve the data acquisition rate by a factor of two. Eventually, the much better power-efficiency, the simplicity and the compactness of the system reported here promises reliable continuous operation without the need for expert intervention during user campaigns.

On the one hand, the demonstrated FWHM pulse duration of 60 fs is clearly shorter than the durations reported from the OPCPA system.[4,15] On the other hand, the main drawback of the nonlinear pulse compression approach is the modest pulse contrast. It is, without further measures, intrinsically limited by the self-phase modulated spectrum as **Figure 2**a implies. The applied nonlinear ellipse rotation method has proven as a viable technique to suppress spurious FEL pump-probe signals originating from laser pulse pedestals. To some extent, pulse energy and duration are compromised by the approach but most user experiments run at low μJ pulse energies, and thus are well compatible with the pulse cleaning method. It is to note that the waveplate settings were found empirically by observing the modulation depth of the output spectrum. Consequently, efficiency as well as contrast could be further improved if necessary by a systematic study of pulse contrast in dependence of polarization ellipticity in the MPC.

The need for only modest pulse energies results in the most compact, cost-efficient and least intensity-noisy near-infrared FEL pump-probe laser among the previously reported ones.[34–36] Only the OPCPA-based laser at LCLS shows comparable long-term pulse energy stability if it is driven with reduced pump power.[34] At full pump power, the LCLS system provides up to 90 W average power after the OPCPA, that is conversion of about 13.5 % of the power provided by the Yb:YAG amplifier. By contrast, the system reported here can exploit about 50 % of the pulse energy available after the chirped pulse amplifier. Only 20 % of the losses stem from the MPC. The main loss source is the flexible grating compressor which could be replaced by efficient dispersive mirrors. Consequently, even compression to the 10 fs-level in a dual stage scheme[31] is expected to be significantly more efficient than the established OPCPA technique. The results presented here highlight the attractiveness of the pulse-compression approach for future FEL pump-probe laser developments. Its scalability to kW average power and pulse energy levels exceeding 100 mJ has just recently been demonstrated.[37,38]

The about 30 fs rms timing jitter may be further reduced with a recently developed nonlinear amplifying loop mirror (NALM) laser oscillator.[39] The jitter refers to the locking of the optical laser to the main laser oscillator. The actual timing instability between FEL and optical pulses may be slightly larger.[20] Pulse arrival time monitoring could be used for full synchronization characterization,[40] is however comparably elaborate for the available XUV to soft-X-ray spectral range.[20] It was shown that the 30 fs rms timing jitter has hardly impact on the temporal resolution since the FEL pulses are comparably long at the monochromator beamline. Therefore, the possible improvements in synchronization and absolute jitter characterization will not lead to significant changes of the temporal resolution.

Whereas the previous OPCPA system was generating pulses at 800 nm central wavelength, the spectra of the pulses from the laser reported here are centered at 1030 nm. As the sample excitation in the user experiments is typically non-resonant, this wavelength shift does not influence the usability of the laser. Recently, an additional harmonic generator delivering light at 517 nm has been implemented (supplement S7). The intensity dependence of the nonlinear frequency conversion additionally cleans the ultrashort pulses as the smooth spectrum of the second harmonic indicates (**Figure S9**b). The third harmonic at 343 nm will be available to the users in the near future which will further extend the applicability of the new laser system.

## 5. Conclusion

A first FEL facility laser was presented which fully relies on the spectral broadening in a multi-pass cell approach. The demonstrated compression factor of 15 to FWHM durations of 60 fs makes it possible to exploit the stability, compactness and efficiency of a high-power Yb:YAG laser emitting ps-level pulses. The short pulse durations in combination with the about 30 fs rms timing jitter are excellently suited for pump-probe experiments at the FLASH PG beamline where temporal resolution is limited by the FEL monochromator. A major challenge of nonlinear pulse compression methods, to generate clean pulses, was tackled by the nonlinear ellipse rotation method which led to a significant reduction of spurious signals in FEL cross-correlation measurements. In conclusion, the new compact and stable facility laser for pump-probe experiments at FLASH meets all major user requirements, promises high reliability during 24/7 operation and will certainly contribute to upcoming cutting-edge ultrafast science experiments at the PG beamlines at DESY's FEL facility.

## 6. Methods

*FROG measurements.* A commercial scanning second-harmonic FROG was used (Mesa Photonics). It contained a 20 μm thin BBO crystal for nonlinear frequency conversion. A beam sampler after the grating compressor was inserted to direct the beam to the measurement device. AOM 3 was used to reduce the number of pulses from 800 to 5 which illuminated the spectrometer in the FROG best. By delaying the trigger of AOM 3 intra-burst delay dependent FROG measurements were done. A 512 x 512 grid was chosen. The FROG errors varied between 0.23 % and 0.68 %. A grating spectrum was measured in parallel by collecting scattered light from the block of the main beam.

$M^2$-*measurements.* A commercial $M^2$-meter was used (Spiricon M200-s) to determine $M^2$. The device uses the 4σ-method, automatically attenuates and calculates the region of interest of the camera image. By means of a beam sampler, light in front of the MPC or after the MPC was steered to the measurement device. The beam camera was triggered and the integration time was set to 40 μs. This allowed to measure $M^2$ at different intra-burst delays.

*Laser stability measurements.* InGaAs photodiodes were used to continuously track laser pulse energies at different parts of the setup. The sampled laser beam was strongly attenuated and focused onto the detector area to reduce beam pointing artifacts. The pulse energies were calibrated by comparing the integrated voltage over one burst with the burst energy measured by a commercial energy meter. From the integrated voltage of a single pulse, the pulse energy was then retrieved. The pulse energies shown in **Figure 3**b were derived from diodes located directly behind the beam stabilization units to minimize fluctuations caused by beam pointing.

*Optical cross-correlator calibration.* To calibrate the optical cross-correlator for drift correction, a delay scan was done (**Figure S6**) and the slope of the balanced photodiode signal with respect to the displacement of the variable translation stage was fitted. For the optical locking of the laser oscillator, the phase of the RF-lock was shifted. The recorded balanced signals were then converted to fs delays.

*Measurements with the HEXTOF instrument at the FLASH1 PG2 beamline.* The monochromatic FEL and optical laser beams impinge on the sample in a collinear configuration thanks to a plan holey mirror in the incoupling section of the beamline, 1.5 m upstream of the sample. The incident angle of the two beams was 68 degree (with respect to sample normal). The focus sizes of the FEL and the optical laser beam were 150 µm × 300 µm and 200 µm × 400 µm, respectively. The FEL photon energy was $h\nu_{FEL} = 112$ eV ($\lambda = 11$ nm). The FEL pulses contained a few thousand photons for core-level spectroscopy.[15] The near-infrared pulse energies were 2.0 µJ, corresponding to 73 GW/cm² peak intensity and 6.4 mJ/cm² peak fluence, respectively. The bursts contained 480 pulses. The photoelectrons were collected by the HEXTOF extractor lens and parallelized onto a multi-channel plate for momentum and delay detection.[15] **Figure 4** shows momentum-integrated data after 30 minutes of averaging. The sample was contaminated with carbon atoms as the usual annealing treatment was skipped for the shown measurement run. The HEXTOF detection scheme allows to collect all electrons emitted above the sample surface giving an effective acceptance angle of $2\pi$ steradians.

*FEL pulse duration estimation.* As the direct measurement of the high energy FEL pulses is difficult, the electron bunches were analyzed in order to estimate the photon pulse durations. A LOLA-type transverse deflecting radio-frequency structure was used the measure the longitudinal electron bunch profile. Its rms duration corresponds roughly to the FWHM of the FEL pulses if the profile is Gaussian.[29] This estimation results in a 120 fs pulse width. The 150 fs duration used in the model is additionally taking into account a 30 fs pulse elongation induced by the plane grating monochromator of the beamline. The elongation was calculated from the XUV beam size on the grating and its line density.[15]

## Acknowledgements

We acknowledge DESY (Hamburg, Germany), a member of the Helmholtz Association HGF, for the provision of experimental facilities. Parts of this research were carried out at FLASH. We thank Holger Meyer and Sven Gieschen from the University of Hamburg for support of the HEXTOF instrument. We acknowledge our colleagues from European XFEL for developing and sharing the drift cross-correlator design.

D:K., M.H., N.W. acknowledge funding by the DFG within the framework of the Collaborative Research Centre SFB 925 - 170620586 (project B2).

# Supporting Information

**Ultrafast MHz-rate burst-mode pump-probe laser for the FLASH FEL facility based on nonlinear compression of ps-level pulses from an Yb-amplifier chain**

*Marcus Seidel, Federico Pressacco, Oender Akcaalan, Thomas Binhammer, John Darvill, Maik Frede, Uwe Grosse-Wortmann, Michael Heber, Christoph M. Heyl, Dmytro Kutnyakhov, Chen Li, Christian Mohr, Jost Müller, Oliver Puncken, Harald Redlin, Nora Schirmel, Sebastian Schulz, Angad Swiderski, Hamed Tavakol, Henrik Tünnermann, Caterina Vidoli, Lukas Wenthaus, Nils Wind, Lutz Winkelmann, Bastian Manschwetus, Ingmar Hartl*

## Supplement S1: Setup Details

### S1.1. Fiber frontend

Ultrafast pulses are generated by a Fabry-Pérot oscillator which comprises a short free-space section with a semiconductor saturable absorber mirror (SAM) for self-starting mode-locked operation, a polarizer and a single mode polarization maintaining (PM) fiber section which includes a Yb-doped gain fiber (Coractive Yb-401 PM) and a 5-nm bandwidth chirped fiber-Bragg-grating (FBG) for dispersion compensation, laser pulse outcoupling and incoupling of 976 nm pump light.[7] The oscillator is operated in the soliton-regime with a net intracavity dispersion of -0.21 ps$^2$. It emits a 54 MHz train of 4 nm bandwidth pulses, centered at 1030 nm with an average power of 7.8 mW. For pump-probe experiments with the FEL, precise synchronization of the oscillator is required. Therefore, two actuators are used for cavity-length control: A > 600 kHz bandwidth piezo-ceramic chip actuator which can translate the SAM with up to 2.2 µm amplitude[41] and a slow piezo-actuated translation stage which can translate the intra-cavity fiber-collimator with up to 120 µm amplitude. The pulses are amplified in three core-pumped single mode PM Yb:fiber amplifiers (YDFA) with 2.8 dB, 10.8 dB and 12.7 dB gain, respectively (**Figure 1**). They are separated by isolators to prevent amplification of backward travelling waves. Behind the first amplifier, the pulses are stretched in 200 m PM fiber to ~35 ps duration for chirped pulse amplification. Behind the second amplifier, a 1 MHz pulse-train is picked. For reducing nonlinearities, the amplifier core-diameter is increased to 12 µm in the YDFA 3. Additionally, a part of the pulse-train is split-off behind the first amplifier and is nonlinearly broadened by self-phase modulation in an additional amplifier (YDFA 4) to generate 180 fs short reference pulses for the balanced cross-correlators in the synchronization section.

*S1.2. Solid-state four-stage amplifier*

The bulk amplifier rods are diode-pumped at 940 nm wavelength. The pump light emerges from multi-mode fibers and is focused into the 7 mm long gain crystals (orange solid lines in **Figure 1**). The laser light enters the crystals from the opposite side. It is matched to the pump light mode in order to efficiently extract the gain and maintain a near Gaussian profile. Pump and laser light are separated by 45° dichroic mirrors. The Yb:YAG 1 amplifier is different from the others in the sense that it is firstly continuously pumped and secondly works in double-pass geometry. A quarter-wave plate rotates the linear input polarization from p-polarization (transverse-magnetic) to s-polarization (transverse-electric) after the beam passes twice. The outgoing pulses are hence reflected by a polarization-sensitive dielectric mirror after amplification to about 5 µJ energy. The beam passes through AOM 2 where a 1.2 ms long laser burst is diffracted into the first order (top panel in **Figure 1**) and sent through amplifiers 2 – 4 for further amplification to about 60 µJ, 130 µJ and 200 µJ on the burst plateau, respectively. Yb:YAG 2 – 4 are only pumped for 1.23 ms at the 10 Hz burst repetition rate. The ms-order pump period sets the gain crystals in a non-stationary temperature state while the laser beam passes. This causes transient thermal lensing and thus imposes challenges to keep the beam parameters constant over the full burst. By imaging the crystal planes into each other, the effect could however be reduced such that the $M^2$-parameter measured behind the grating compressor hardly varies over the burst (Supplement S2). It is always about 1.1. Nevertheless, a relative waist area variation of 30 % and a focal plane shift of 64 % relative to the mean Rayleigh range are measured inside the $M^2$-meter. This however has only little influence on the pulse compression performance over the burst as shown in section 2.2.

AOM 3 is used to shape the amplified laser bursts on a single pulse basis. Diffraction efficiencies of 86 % into the first order are routinely achieved. For burst flattening, the energies of the diffracted pulses are measured by a photodiode. An error signal is generated from the energy difference to the set value. A control loop adjusts the voltage applied to the AOM 3 modulation port until the error converges to zero for the individual pulses. Typically, the feedback loop flattens the burst to pulse-to-pulse energy fluctuations of 3 % rms. This is limited by electronic noise in the photodiode readout. The AOM can in-principle also serve as attenuator or tailor arbitrary burst shapes. This is however rather relevant in the similar photocathode laser setup,[8] and thus the AOM 3 control loop is typically set to produce a flat burst with up to 800 pulses of about 150 µJ energy. This corresponds to 150 W intra-burst average power if the system runs at 1 MHz repetition rate. Behind the amplifier, the stretched pulses are compressed again in a single grating four-pass compressor.

*S1.3. Multi-pass cell*

Three 1 mm thin anti-reflection (AR) coated silica plates with 3 cm spacing were placed in the center of an about 350 mm long MPC consisting of two spherical mirrors with 200 mm radius of curvature. The spot size in the center of the cell was adjusted with a telescope of two best form convex lenses such that curvature of a Gaussian beam would match the curvature of the spherical mirrors. Kerr lensing of the silica plates was not considered for mode-matching since the intensity of the incoming beam was about a factor six lower than the damage threshold of the AR-coating, i.e. a small effective mode-mismatch was tolerated by the setup. The beam entered and exited the cell after 31 roundtrips by means of a pick-off mirror with about 4 mm height. After passing a half-wave plate and a polarizing beam splitter for polarization cleaning, the spectrally broadened pulses were compressed to about 60 fs. The MPC transmission is about 80 % although the beam passes 372 AR-coated surfaces. For pre-chirped pulses yielding a reduced Fourier transform limit of 80 fs after the MPC, 90 % transmission could be reached. If quarter-wave plates were used for pulse contrast improvements, the transmission strongly depended on the waveplate settings.

**Supplement S2: M² measurements**

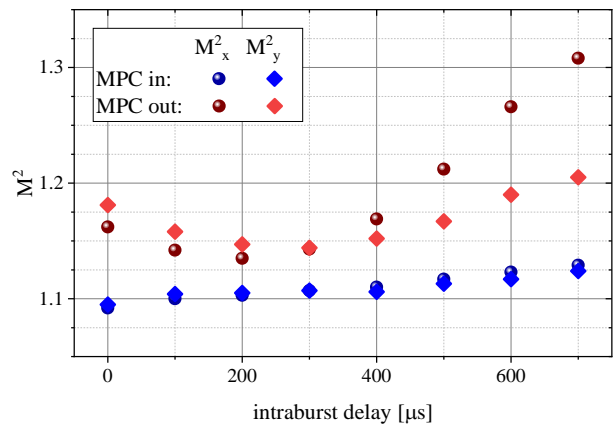

**Figure S1.** M²-parameters measured for different trigger delays of the M²-meter camera within the burst. The measurements were taken in front of and behind the MPC.

To measure the beam parameter M² at different instances of the burst. The CCD camera of M²-meter was triggered in 100 µs delay steps and the integration time was set to its minimum of 40 µs. The timing relative to the burst was retrieved by reducing the pulse number in the burst to 40 by AOM 3 and maximizing the signal of the CCD camera by shifting its trigger delay. A beam sampler was inserted in front of and behind the MPC to measure $M^2$ at both locations. The results are shown in **Figure S1**. The $M^2$-values after the first compressor are all about 1.1 in horizontal (x-) and vertical (y-) direction. The $M^2$-values behind the MPC are slightly higher. They are between 1.1 and 1.2 except from the end of the burst where the $M^2$ in horizontal direction increases to 1.3.

**Supplement S3: Spatial homogeneity after nonlinear broadening in the multi-pass cell**

To check if the center and the wings of the beam are equally spectrally broadened, a 4f-imaging spectrograph was used. The grating of the device dispersed the beam filtered by a narrow entrance slit in x-direction. The y-direction displays the one-dimensional beam profile. To investigate spatial chirp that can be introduced by the grating compressors in x-direction, a periscope was used to flip the horizontal and the vertical beam axes. Consequently, the y-axis of the CCD camera showed the horizontal axis of the laser beam emerging from the second grating compressor. **Figure S2** shows that the beam was homogeneously broadened in the MPC and that it is not spatially chirped.

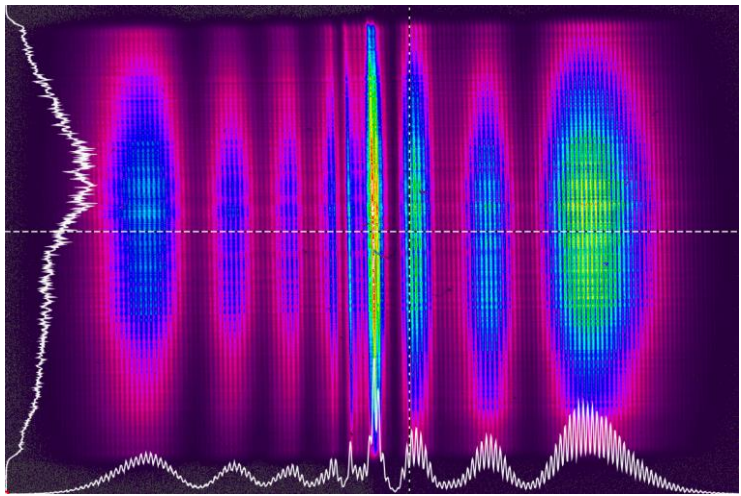

**Figure S2.** Measured spectra (x-direction) across the horizontal axis of the beam emerging from the grating compressor. It is slightly cut by the entrance slit of the spectrometer. The shape of the spectra looks very homogeneous across the beam in good agreement with previous MPC studies. Also, the introduction of spatial chirp by the grating compressor could be minimized. The fast modulations on the spectrum were not observed anywhere else in the setup and must originate from the beam sampler used to direct a portion of the main beam to the imaging spectrograph.

**Supplement S4: Laser long-term stability**

*S4.1. Beam position and pointing*

The beam position and pointing data logged over one user-experiment day is shown in supplementary **Figure S3**. The position of the beam on the camera was calculated by the first moment of the CCD counts. The far-field image is generated by a 100 mm focal length lens transforming the beam from real to wavevector space.

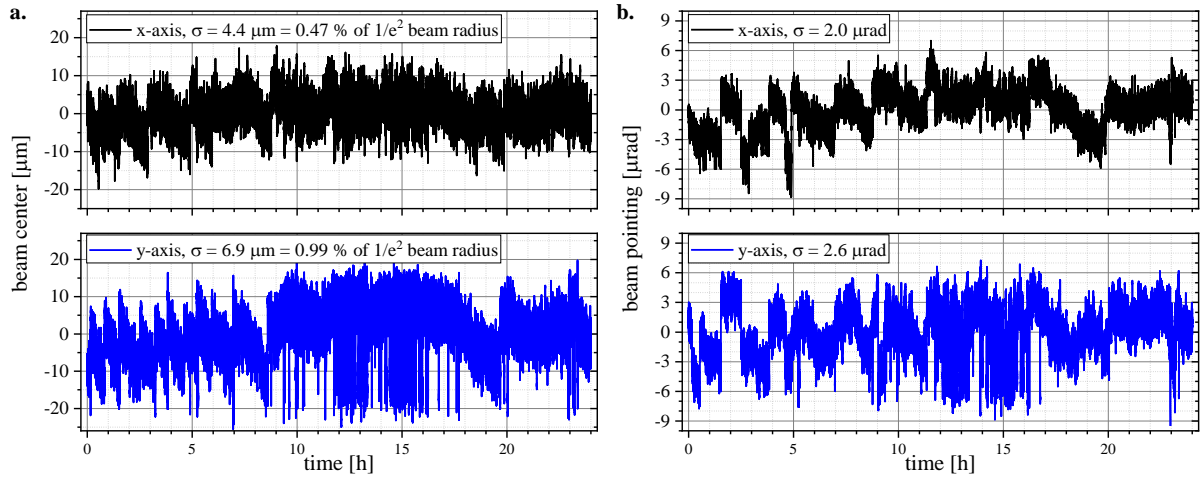

**Figure S3. a.** Position of beam center relative to mean position over the 24-hour measurement period for the x- (top) and the y-axis (bottom). **b.** Beam propagation direction relative to mean direction over 24-hour measurement period for the x- (top) and the y-direction (bottom). The direction was calculated by $\theta_{x,y} = \tan\frac{\Delta(x,y)}{f}$ where f = 100 mm and $\Delta(x,y)$ were the position deviations from mean on the far-field camera. The jumps of beam center and direction, respectively, were caused by the beam stabilization units.

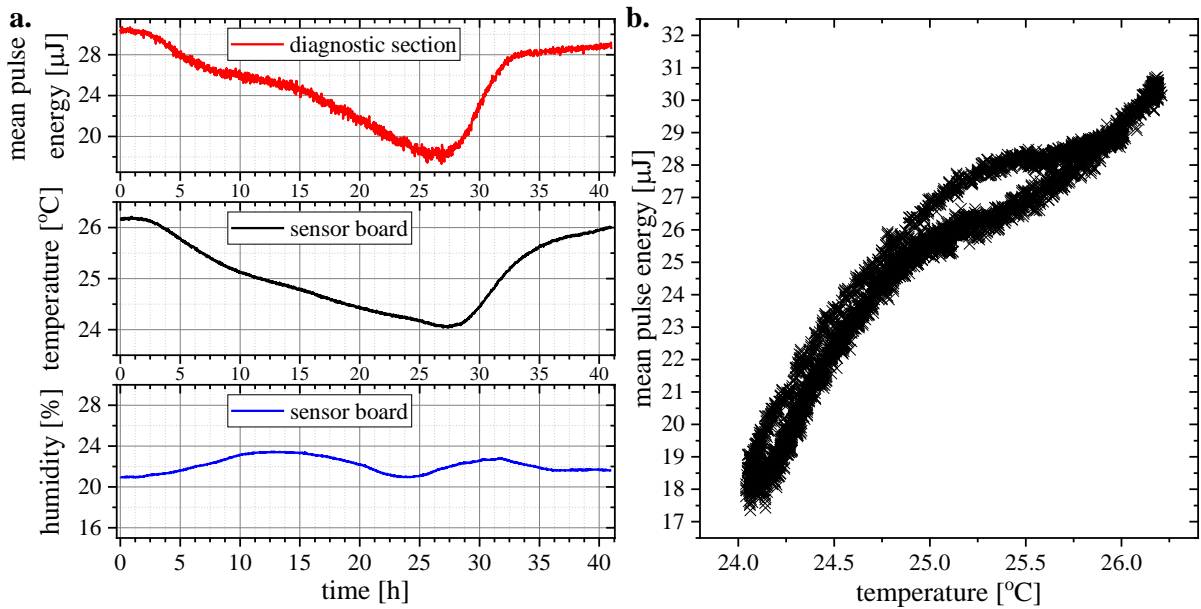

**Figure S4. a.** Logged data of mean pulse energy, temperature measured at a sensor board on the optical table of the laser and humidity. Due to partial failure of air conditioning in the FLASH hall, the temperature changed by more than 2 °C. The humidity change was moderate during this time. **b.** Correlation between measure temperature and measured mean pulse energy. The best linear fit results in a slope of 5.2 µJ/°C corresponding to 17 % of initial pulse energy per °C.

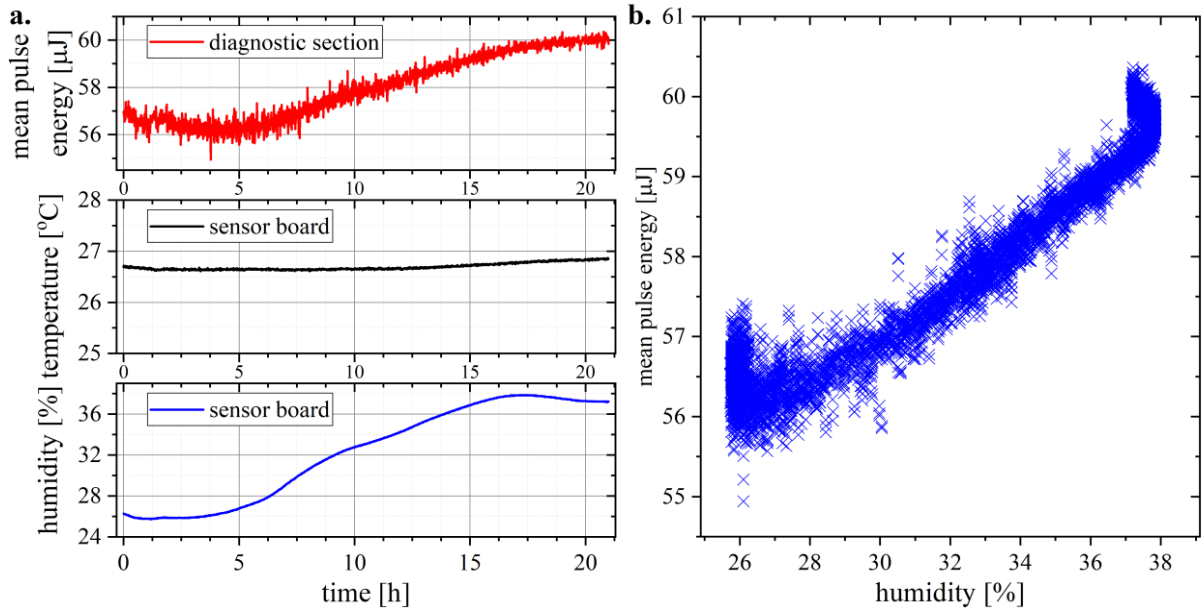

**Figure S5. a.** Logged data of mean pulse energy, temperature measured at a sensor board on the optical table of the laser and humidity. Whereas the temperature was relatively stable during the observation time, the humidity changed by about 12 % and the mean pulse energy by about 7 %. **b.** Correlation between measure temperature and measured mean pulse energy.

### S4.2. Correlations between temperature, humidity and mean pulse energy

To derive the sensitivity of the mean pulse energy at the diagnostics section to temperature and humidity changes, respectively, logged data from first, a period when an air conditioning failure at the FLASH1 hall occurred, causing an about 2 °C temperature change in the modular delivery station (**Figure S**), and second, when humidity changed from about 26 % to 38 % during half a day (**Figure S**) was analyzed. These strong changes dominated the burst energy variation measured at the diagnostic section, and thus a clear correlation between the quantities became visible. The correlation between temperature and pulse energy is with about 5 μJ/°C particularly strong. It is to note that no significant change in spectral bandwidth was observed although the measured mean pulse energy was reduced by 40 %. Consequently, the energy drops most likely behind the MPC. The exact origin is however yet unknown. In addition to the change of energy, a strong drift was measured by the optical drift correlator. The best linear fit of the correlation in **Figure S**a results in a slope of 0.3 μJ/% humidity, corresponding to 0.5 % of the maximum pulse energy per % humidity.

## Supplement S5: Optical cross-correlators

The balanced optical cross-correlator for the locking of the oscillator is a type I two-color ($\lambda_1$ = 1030 nm, $\lambda_2$ = 1550 nm) cross-correlator, whereas the one used for drift stabilization of the laser system uses type II phase-matching and the center wavelengths of both pulses are identical (1030 nm). In both correlators reference and signal are split each into two replicas and overlapped at slightly different delays. The sum frequencies of both signal-reference-pairs are generated and their differential photocurrent is measured by balanced detectors generating the error signal for the feedback loops. Locking of the oscillator to the facility reference is done by an FPGA based controller whereas a slower MATLAB script suffices for the drift compensating feedback. The signal is relatively insensitive to amplitude fluctuations and exhibits an almost linear slope near time zero. **Figure S6** shows an example of a calibration scan of the drift correlator. The fast optical locking method is described in more detail in the references.[20,42]

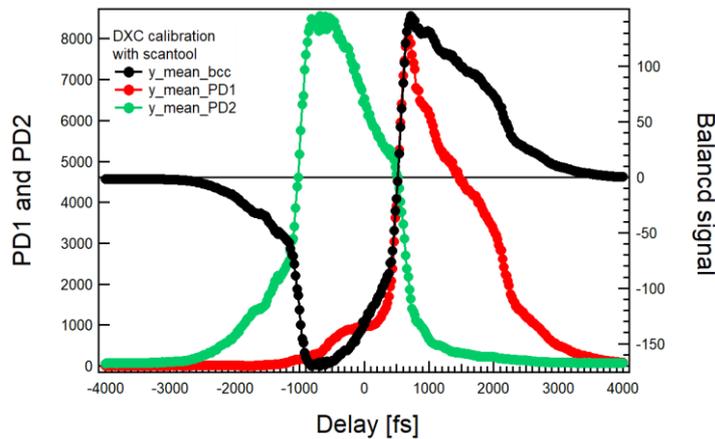

**Figure S6.** Exemplary calibration scan of the drift correlator: For negative delays, only the photodiode PD2 gives a signal, resulting in negative balanced signal $y_{bcc} = y_{PD1} - y_{PD2}$. Vice versa, for positive delays, only at PD1 a photocurrent is generated, yielding a positive balanced signal. In between, both diodes contribute to $y_{bcc}$ resulting in a zero-crossing with an adjacent linear slope of about 220 fs range. This is used as error signal for the feedback loop.

## Supplement S6: FEL cross-correlation measurements

To demonstrate the impact of nonlinear polarization ellipse rotation on the optical pulse-FEL-cross-correlation, a time-of-flight measurement with the pulse shown in **Figure 2**a was recorded. The FEL was tuned to $h\nu_{FEL}$ = 256 eV ($\lambda$ = 4.8 nm) in this experiment. By the procedure described in the methods section, the FEL pulses were estimated to be Gaussian with 120 fs FWHM duration. In this case the estimation was well justified because the longitudinal electron bunch structure was also Gaussian. The energy of the optical pulses was set to $\bar{E}_p$ = 4.3 μJ which resulted in a peak fluence of 13.7 mJ/cm$^2$ and a peak irradiance of about 160 GW/cm$^2$ being 2.25 times higher than in the measurement shown in **Figure 4**. Furthermore, the W(110) sample was annealed before the measurement and the averaging time of the measurement was 90 minutes. This resulted in a clearly less noisy binding energy–pulse delay–map (**Figure S**a) but also in a highly nonlinear sample response. The delay-integrated data in **Figure S**a clearly displays transient states dressed with up to 4 near-infrared photons (at $E_{b,4h\nu_{nIR}} \approx -26.4$ eV). Consequently, the energy-integrated data around the single-photon dressed state appears strongly power-broadened (**Figure S**b) which leads to 30 % amplitude of the satellite pulse around 900 fs delay with respect to the main peak.

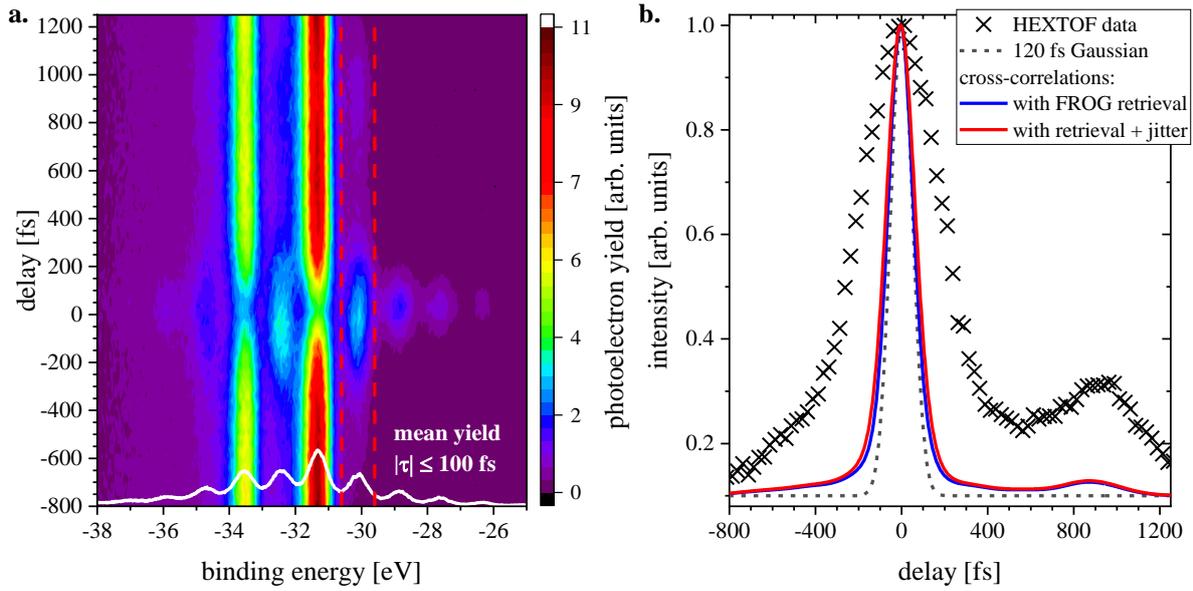

**Figure S7**. Measured photoelectron spectra of the W(110) 4f core level as function of time delay between optical laser and FEL ($h\nu_{FEL} = 256$ eV $\approx 213 \times h\nu_{nIR}$). The white line shows the integrated photoelectron yield in the $\pm 100$ fs delay range. Due to the high pump power, optical photon dressed states are visible up to the fourth order near -26.4 eV. The yield around the most prominent single-photon side-band at $E_{b,1} + h\nu_{nIR} \approx -30.1$ eV was integrated in a range of 1.0 eV (red dashed lines) and is shown in **b**. The experimental data is represented by the black crosses. At 900 fs delay, strong photo-emission intensity reveals the presence of laser post-pulses. The delay is in good agreement with the modeled cross-correlation. The amplitude is however clearly larger. This is mainly attributed to the higher-order side-bands that involve multiple near-infrared photons and subsequently deplete the single-photon side-band. The cross-correlation was modeled with Eq. (1) by means of the FROG retrieval from **Figure 2**a. The blue (red) solid line shows the expected cross-correlation excluding (including) 30 fs rms timing jitter. The estimated FEL pulse is represented by the grey dashed line.

A better agreement between cross-correlation model (Eq. (1) main text) and HEXTOF measurement is attained by rescaling the measurement data by a factor 1/8 (**Figure S**). It becomes clear that the optical satellite pulse retrieved by the FROG measurement reproduces itself in the FEL cross-correlation and may thus complicate the analysis of samples with a non-instantaneous response on the 1 ps-order.

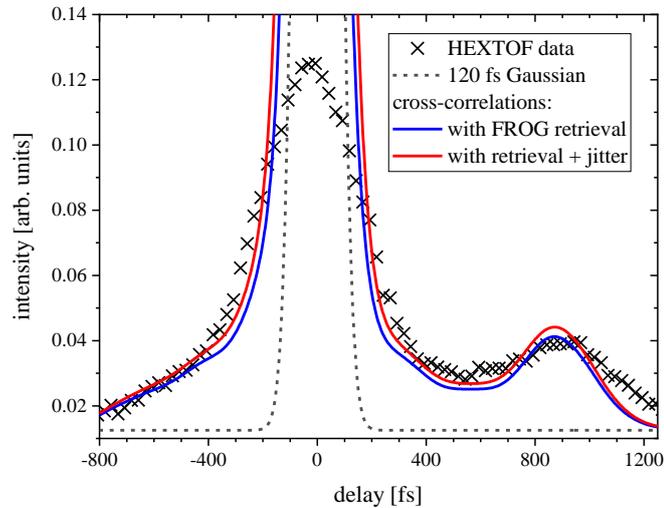

**Figure S8.** FEL cross-correlation measurement shown in **Figure S7**b with experimental data rescaled by 1/8 to emphasize the presence of a post-pulse at 900 fs delay in both the FROG and the FEL cross-correlation measurement.

## Supplement S7: Second harmonic generation

A stage for second harmonic generation (SHG) was implemented between the laser diagnostics section (D) and the laser delivery section (E) of **Figure 1**. The fundamental is focused into a 2 mm thick LBO crystal which converts 42 % of the incident power to green (**Figure S9**a). The second harmonic spectrum is smooth (**Figure S9**b). This results in an about 63 fs long transform-limited pulse without a significant pedestal structure (**Figure S9**c). The implementation of beam and pulse diagnostics for the second harmonic is yet ongoing.

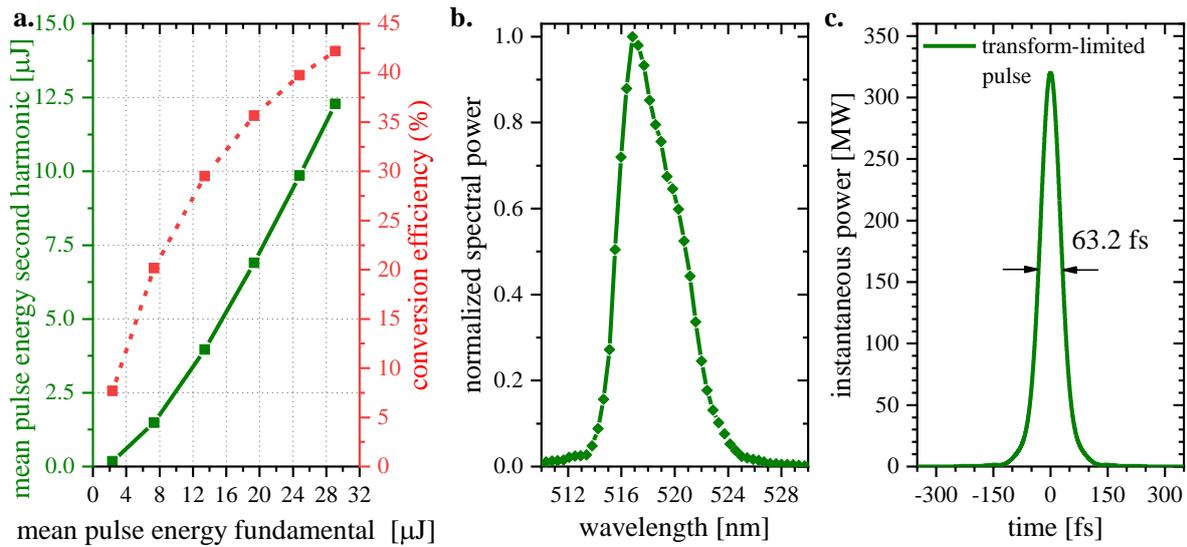

**Figure S9**. **a.** Second harmonic mean pulse energy in dependence of fundamental pump energy (green solid line) and corresponding power conversion efficiency (red dashed line). Nonlinear polarization ellipse rotation was used to clean the fundamental pulse. Second harmonic was generated in a 2 mm thick LBO crystal. **b.** Spectrum of the second harmonic at maximum pulse energy (12 μJ) and **c.** transform-limited pulse derived from Fourier transformation of the spectrum. Contrary to the fundamental spectrum (**Figure 2**b), the spectrum of the SH is not modulated and thus the transform-limited pulse of the SH does not show distinct pedestals like the transform-limited pulse of the fundamental spectrum in **Figure 2**a.